\documentclass[twocolumn,showpacs,preprintnumbers,superscriptaddress,amsmath,amssymb]{revtex4}
\usepackage{graphicx}
\usepackage{dcolumn}
\usepackage{bm}
\usepackage{amssymb}
\usepackage{amssymb}
\def\>{\rangle}

\begin{document}

\title{Transition from weak to strong measurements by nonlinear quantum feedback control}

\author{Jing Zhang}\email{jing-zhang@mail.tsinghua.edu.cn}
\affiliation{Department of Automation, Tsinghua University,
Beijing 100084, P. R. China} \affiliation{Center for Quantum
Information Science and Technology, Tsinghua National Laboratory
for Information Science and Technology, Beijing 100084, P. R.
China} \affiliation{Department of Physics and National Center for
Theoretical Sciences, National Cheng Kung University, Tainan
70101, Taiwan}
\author{Yu-xi Liu}
\affiliation{Institute of Microelectronics, Tsinghua University,
Beijing 100084, P. R. China}
\affiliation{Center for Quantum
Information Science and Technology, Tsinghua National Laboratory
for Information Science and Technology, Beijing 100084, P. R.
China}
\author{Re-Bing Wu}
\affiliation{Department of Automation, Tsinghua University,
Beijing 100084, P. R. China}
\affiliation{Center for Quantum
Information Science and Technology, Tsinghua National Laboratory
for Information Science and Technology, Beijing 100084, P. R.
China}
\author{Chun-Wen Li}
\affiliation{Department of Automation, Tsinghua University,
Beijing 100084, P. R. China}
\affiliation{Center for Quantum
Information Science and Technology, Tsinghua National Laboratory
for Information Science and Technology, Beijing 100084, P. R.
China}
\author{Tzyh-Jong Tarn}
\affiliation{Department of Electrical and Systems Engineering,
Washington University, St.~Louis, MO 63130, USA}
\affiliation{Center for Quantum Information Science and
Technology, Tsinghua National Laboratory for Information Science
and Technology, Beijing 100084, P. R. China}

\date{\today}

\begin{abstract}
We find that feedback control may induce ``pseudo" nonlinear
dynamics in a damped harmonic oscillator, whose centroid
trajectory in the phase space behaves like a classical nonlinear
system. Thus, similar to nonlinear amplifiers (e.g., rf-driven
Josephson junctions), feedback control on the harmonic oscillator
can induce nonlinear bifurcation, which can be used to amplify
small signals and further to measure quantum states of qubits.
Using the cavity-QED and the circuit QED systems as examples, we
show how to apply our method to measure the states of two-level
atoms and superconducting charge qubits.
\end{abstract}


\maketitle

\section{Introduction}\label{s1}

Quantum feedback control~\cite{HMWisemanbook} is one of the
central parts of quantum control theory and
applications~\cite{Alessandro,HMabuchi and
NKhaneja,DYDong,NGanesan,Zhang2} owing to its potential ability to
improve the stability and robustness of the system. Besides the
extensive theoretical
studies~\cite{Belavkin,Wiseman1,Doherty1,James,Korotkov}, recent
rapid developments on sensitive measurements and manipulation
techniques in atom-optical~\cite{Mabuchi,Cook,Gillett} and
solid-state systems~\cite{You} have made it possible to implement
quantum feedback control in laboratories.

In atom-optical systems, atomic ensembles are put in high-Q
optical cavities, so that the information of the states of the
atomic ensembles can be extracted by the probe lights transmitted
through the optical cavities. Photocurrents induced by the probe
lights are processed by field programmable gate arrays to generate
real-time control signals, which can be fed back to design the
electromagnetic fields imposed on the atomic
ensembles~\cite{Mabuchi}. Possible applications of quantum
feedback control in such systems include state
stabilization~\cite{State stabilization}, entanglement
production~\cite{Entanglement production}, spin
squeezing~\cite{Spin squeezing}, and state
discrimination~\cite{Cook,Gillett}. Similar studies can be found
in solid-state systems~\cite{Ouyang}, e.g., the superconducting
circuit-QED systems, in which quantum feedback control has been
proposed to cool and squeeze the motion of a nanomechanical
resonator~\cite{Ruskov1,Hopkins1,Zhang1,Jahne}. In these studies,
the position of the nanomechanical resonator can be measured by a
single electron transistor~\cite{Hopkins1} or a
rf-SQUID~\cite{Zhang1,Buks,Etaki}.

Most existing theoretical studies on quantum feedback control are
concentrated on linear quantum systems, i.e., the dynamical
equations of the system are linear in the Heisenberg picture and
the feedback controls are linear functions of the system states,
which can often be reduced to standard classical control problems,
e.g., the Linear Quadratic Gaussian control
problem~\cite{Doherty1,James}. However, essential differences
arise when we study nonlinear quantum systems: (i) linear quantum
systems possess evenly spaced discrete energy spectra, while the
distributions of energy levels in nonlinear systems are uneven or
continuous; (ii) linear quantum system starting from a Gaussian
state always stays in a Gaussian state during the evolution, while
nonlinear quantum dynamics generally distorts the wavepacket.
Moreover, the presence of the relaxation and dephasing effects in
nonlinear quantum systems may give rise to inherent phenomena in
nonlinear classical systems, e.g., chaos~\cite{Habib} and
bifurcation~\cite{Armen}. These dissipation-induced nonlinear
effects~\cite{Habib,Armen} have various applications in
laboratories. For example, nonlinear dynamical bifurcation of a
rf-driven Josephson junction is used to amplify small
signals~\cite{Siddiqi1}, and further applied to the readouts of
the superconducting qubit states in
experiments~\cite{Siddiqi1,Lupascu}.

Here we propose a method to mimic nonlinear dynamics using a
harmonic oscillator with the feedback control. Such a proposal can
be widely applied, especially in circumstances where a nonlinear
amplifier like the rf-driven Josephson junction is not achievable.
It should be noticed that the manipulation of nonlinear effects
via feedback has been widely studied in classical
systems~\cite{Isidori}. However, to what extent can we apply this
to quantum systems? In particular, an extensively discussed
question is {\it whether or not the nonlinear
effects{\rm~\cite{Jacobs}} can be produced in linear quantum
systems by quantum feedback control.} Here we address this
question by examining a damped harmonic oscillator driven by a
nonlinear feedback control, and shows that using this ``pseudo"
nonlinear amplifier can read out quantum states of qubits.

Our proposal is motivated by the recent developments for the
quantum state readouts in superconducting quantum circuits via a
nonlinear amplifier~\cite{Siddiqi1} and the quantum information
processing using the cavity QED effect (e.g., in atom-optical
systems, cavity quantum-dot systems, and the systems for the
interaction between superconducting qubits and the transmission
line resonator). In contrast to the measurement of the quantum
states using a nonlinear amplifier (e.g., to read out the states
of a superconducting qubit using a rf-driven Josephson
junction~\cite{Siddiqi1}), one merit for our study is that in the
long-time limit we can analytically analyze the dynamics of the
qubit-oscillator system even in the ``nonlinear" regime, which
makes it possible to see how to control the rate of the
information extraction so as to balance between the measurement
sensitivity and the measurement-induced disturbance in the
bifurcation readout regime. Our study can be applied to the
measurement of the states of an atom inside a cavity, quantum dots
interacting with single-mode cavity field, or superconducting
qubits in the circuit QED systems. Without loss of generality,
below we choose the atom-cavity system and the circuit QED system
as examples to demonstrate our method, because these two kinds of
systems are more experimentally controllable and developed very
quickly in recent years. Thus our proposal might be more possible
to be demonstrated in these systems.

The paper is organized as follows: in Sec.~\ref{s2}, we present
our feedback control proposal by a general model in which a damped
harmonic oscillator is measured by a homodyne detection and then
driven by the output field of the feedback control loop. The model
of a system, in which a harmonic oscillator is dispersively
coupled to a qubit and driven by an outer feedback control
circuit, is discussed in Sec.~\ref{s3}. Also, the analysis of the
bifurcation-induced qubit readout by a harmonic oscillator with
``pseudo" nonlinear dynamics is given in this section. The
applications to the atom-cavity systems and the circuit QED
systems are shown in Sec.~\ref{s5}, and the conclusions and
discussions are given in Sec.~\ref{s6}.

\section{Controlled damped harmonic oscillator}\label{s2}

Consider a damped harmonic oscillator with an angular frequency
$\omega$ and a damping rate $\gamma$, which is driven by a control
signal $u_t$. Let us assume that the decay of the harmonic
oscillator can be detected by a homodyne measurement with
efficiency $\eta$. Under the weak measurement
assumption~\cite{Sahel}, the dynamics of the harmonic oscillator
with the state $\rho$ and the measurement output $dy$ for
measuring the position operator $x$ can be described by the
stochastic
equations~\cite{Wiseman1,Doherty1,James,Korotkov,Mabuchi}:
\begin{eqnarray}
\label{General stochastic master equation}
d\rho&=&-\frac{i}{\hbar}[H,\rho]dt+\gamma\mathcal{D}[a]\rho
dt+\sqrt{\eta\gamma}\mathcal{H}[a]\rho dW,\\
\label{General measurement output}dy&=&\langle x\rangle
dt+\frac{1}{\sqrt{2\eta\gamma}}dW,
\end{eqnarray}
where $\langle x\rangle={\rm tr}\left(x\rho\right)$ is the average
of the position operator $x$; $a$ and $a^{\dagger}$ are the
annihilation and creation operators of the driven harmonic
oscillator whose Hamiltonian is
\begin{equation}\label{Hamiltonian of the controlled harmonic oscillator}
H=\hbar\omega a^{\dagger}a+\hbar u_t\, x.
\end{equation}
The term $\hbar u_t\, x$ represents the interaction between the
harmonic oscillator and a time-dependent classical control field
characterized by $u_t$. The superoperators $\mathcal{D}[a]\rho$
and $\mathcal{H}[a]\rho$ are defined by:
\begin{eqnarray*}
\mathcal{D}[a]\rho&=&a\rho
a^{\dagger}-\frac{1}{2}\left(a^{\dagger}a\rho+\rho
a^{\dagger}a\right),\\
\mathcal{H}[a]\rho&=&a\rho+\rho a^{\dagger}-{\rm
tr}\left(a\rho+\rho a^{\dagger}\right)\rho.
\end{eqnarray*}
The measurement is performed over the position operator
\begin{eqnarray*}
x=\frac{1}{\sqrt{2}}\left(a+a^{\dagger}\right),
\end{eqnarray*}
whose conjugate momentum operator is
\begin{eqnarray*}
p=\frac{i}{\sqrt{2}}\left(a^{\dagger}-a\right).
\end{eqnarray*}
$dW$ is a
measurement-induced Wiener noise satisfying
\begin{eqnarray*}
E(dW)=0,\quad(dW)^2=dt
\end{eqnarray*}
with $E(\cdot)$ representing
the ensemble average over the stochastic noise.

As shown in Eq.~(\ref{General measurement output}), $dW$
represents the measurement-induced noise in the measurement output
of the homodyne detection. To reduce the influence of the noise in
the feedback control design, we take the time average
\begin{equation}\label{Time-averaged measurement output}
Y_t=\frac{1}{t}\int_0^t\left(\langle x\rangle
d\tau+\frac{1}{\sqrt{2\eta\gamma}}dW\right)=\frac{1}{t}\int_0^t\,dy
\end{equation}
of the output signal, where $t\gg1/\gamma$, as an estimation of
the position $x$. Taking the long-time average of the stochastic
signals is an effective filtering strategy to extract stationary
signals from the background noises, and the time-average signal,
e.g., $Y_t$ in Eq.~(\ref{Time-averaged measurement output}), can
be further used to design feedback control. Such a control design
has been used in the literature to prepare desired quantum states
(see, e.g., Ref.~\cite{JKStockton} for the Dicke state
preparation).

With these considerations, we apply the following nonlinear
feedback control
\begin{equation}\label{Three-order nonlinear feedback control}
u_t=f\left(Y_t\right)=-k_1 Y_t+k_3 Y_t^3-k_0,
\end{equation}
to the original system (\ref{General stochastic master equation})
described by the Hamiltonian in Eq.~(\ref{Hamiltonian of the
controlled harmonic oscillator}), where the positive numbers
$k_0,\,k_1,\,k_3$ are the control parameters. In the following
discussions, the control parameters $k_1$ will be chosen such that
$k_1>\gamma$.

The nonlinear feedback control given by Eq.~(\ref{Three-order
nonlinear feedback control}) induces a pitchfork static
bifurcation by varying the bifurcation parameter $\omega$ near the
bifurcation point:
\begin{equation}\label{Critical angular frequency for static bifurcation}
\omega^*=\frac{1}{2}\left(k_1-\sqrt{k_1^2-\gamma^2}\right).
\end{equation}
In fact, if the initial state of the harmonic oscillator is a
Gaussian state $\rho_0$ with
\begin{eqnarray}\label{First and second-order quadratures of the initial Gaussian state}
&\langle x \rangle_{\rho_0}=0,\quad \langle p
\rangle_{\rho_0}=0,&\nonumber\\
&\langle x^2 \rangle_{\rho_0}-\langle x
\rangle_{\rho_0}^2=V_{x_0},&\nonumber\\
&\langle p^2
\rangle_{\rho_0}-\langle p \rangle_{\rho_0}^2=V_{p_0},&\nonumber\\
&\frac{1}{2}\left\langle px+xp\right\rangle_{\rho_0}-\langle x
\rangle_{\rho_0}\langle p \rangle_{\rho_0}=C_{x_0p_0},&
\end{eqnarray}
and if the bifurcation parameter $\omega$ is below $\omega^*$,
then the state of the harmonic oscillator given by
Eq.~(\ref{General stochastic master equation}) converges to the
following stationary coherent state
\begin{equation}\label{alpha0infinity}
\left|\alpha_0^{\infty}\right\rangle=\left|\frac{1}{\sqrt{2}}\left(x_0^{\infty}+i
p_0^{\infty}\right)\right\rangle,
\end{equation}
where
\begin{equation}\label{x0infinity}
x_0^{\infty}=\frac{2\omega}{\gamma}p_0^{\infty}=k_0\frac{\omega}{\omega^2-k_1\omega+\gamma^2/4}.
\end{equation}
However, if the bifurcation parameter $\omega$ exceeds $\omega^*$,
the original stationary state given by Eq.~(\ref{alpha0infinity})
becomes unstable, and two new branches of stationary coherent
states appear
\begin{equation}\label{alpha12infinity}
\left|\alpha_{1,2}^{\infty}\right\rangle=\left|\frac{1}{\sqrt{2}}\left(x_{1,2}^{\infty}+i
p_{1,2}^{\infty}\right)\right\rangle,
\end{equation}
where
\begin{equation}\label{x12infinity}
x_{1,2}^{\infty}=\frac{2\omega}{\gamma}p_{1,2}^{\infty}=\pm\sqrt{\frac{-\omega^2+k_1\omega-\gamma^2/4}{k_3\omega}}.
\end{equation}
The analyses of the above results can be found in Appendix
\ref{Dynamics of the Controlled Harmonic Oscillator}. {\it It can
be verified that $\left|\alpha_{1,2}^{\infty}\right\rangle$ are
far away from the original stationary state
$\left|\alpha_0^{\infty}\right\rangle$ if the parameters
$k_0,\,k_3$ are small enough such that:}
\begin{equation}\label{k0 and k3}
k_0,\,k_3\ll\left|\frac{\omega^2-k_1\omega+\gamma^2/4}{\omega}\right|.
\end{equation}

To give more insights about the above results, we see that $Y_t$
and $\bar{x}=E\left(\langle x \rangle\right)$ coincide together in
the long-time limit (see Eqs.~(\ref{Long-time limit of the
first-order quadratures}) and (\ref{vec(x12)}) in Appendix
\ref{Dynamics of the Controlled Harmonic Oscillator}), and thus
$\left(x_0^{\infty},\,p_0^{\infty}\right)^T$ and
$\left(x_{1,2}^{\infty},\,p_{1,2}^{\infty}\right)^T$ are just the
stationary states of the dynamical equation:
\begin{eqnarray}\label{Dynamical equation with three-order nonlinear terms}
\dot{x}&=&-\frac{\gamma}{2}x+\omega p,\nonumber\\
\dot{p}&=&-\left(\omega-k_1\right)x-k_3 x^3-\frac{\gamma}{2}p+k_0.
\end{eqnarray}
In fact, it can be verified that the above equation has one stable
equilibrium $\left(x_0^{\infty},\,p_0^{\infty}\right)^T$ when
$\omega<\omega^*$ and two stable equilibria
$\left(x_{1,2}^{\infty},\,p_{1,2}^{\infty}\right)^T$ when
$\omega>\omega^*$, which coincides with the results given in
Eqs.~(\ref{x0infinity}) and (\ref{x12infinity}).

It should be pointed out that the feedback control presented here
is different from the Markovian feedback~\cite{Wiseman1} and Bayes
feedback controls~\cite{Doherty1} in which the white noise term
$dW$ in Eq.~(\ref{General measurement output}) is looked as an
innovation information term obtained by measuring a single system
and used to update the feedback control. In contrast, our proposal
is quite similar to a feedback control proposal based on the
measurement over an ensemble of harmonic oscillators. In fact, the
measurement output given by Eq.~(\ref{General measurement output})
can be reexpressed as:
\begin{eqnarray*}
I(t)=\langle x \rangle+\frac{1}{\sqrt{2\eta\gamma}}\xi(t),
\end{eqnarray*}
where $\xi(t)$ satisfies
\begin{eqnarray*}
E\left(\xi(t)\right)=0,\quad \xi(t)\xi(t')=\delta(t-t').
\end{eqnarray*}
Then, we take the average of the measurement output over the
quantum ensemble to obtain
$\bar{x}=E\left(I(t)\right)=E\left(\langle x \rangle\right)$ which
can be further used to design feedback control
$u_t=u\left(\bar{x}\right)$, and the same feedback control is
imposed on each system in the quantum ensemble. Here, we have only
a single system, and thus, different from the above ensemble
feedback control proposal, we take the time-average $Y_t$ given in
Eq.~(\ref{Time-averaged measurement output}) to replace the
ensemble average $\bar{x}$. Since the system state given by
Eq.~(\ref{General stochastic master equation}) tends to a
stationary state given by Eq.~(\ref{alpha0infinity}) or
(\ref{alpha12infinity}), $Y_t$ and $\bar{x}$ coincide together in
the long-time limit from the ergodic theory~\cite{Peterse} (see
Eqs.~(\ref{Long-time limit of the first-order quadratures}) and
(\ref{vec(x12)}) in Appendix \ref{Dynamics of the Controlled
Harmonic Oscillator}). Thus, our feedback control proposal is
similar to the ensemble nonlinear feedback control in the
long-time limit. Compared with the existing feedback control
proposals, e.g., the Bayes feedback, our method is more robust to
the uncertainty of the initial state. In fact, the same stationary
states given by Eqs.~(\ref{alpha0infinity}) and
(\ref{alpha12infinity}) can be obtained if the initial state of
the system deviates slightly from the Gaussian state given by
Eq.~(\ref{First and second-order quadratures of the initial
Gaussian state}).

To show the validity of our feedback control proposal, let us see
some numerical examples. By setting the system parameters as:
$\gamma=250$ MHz, $k_0=50$ MHz, $k_1=500$ MHz, $k_3=50$ MHz,
$\omega=3.5$ MHz, $x_0=p_0=0$, $V_{x_0}=V_{p_0}=1$, and $C_{x_0
p_0}=0$, the simulation results in Fig.~\ref{Fig of the pitchfork
bifurcation}a show that both $\langle x \rangle$ and $Y_t$
converge to a stationary state $x_0^{\infty}=0.0126\approx 0$
given by Eq.~(\ref{x0infinity}). However, if we tune the
oscillating frequency of the harmonic oscillator such that
$\omega=65$ MHz, the stochastic trajectories of $\langle x
\rangle$ and $Y_t$ are separated into two branches which tend to
two different stationary states $x_{1,2}^{\infty}=\pm1.95$ given
by Eq.~(\ref{x12infinity}). The above simulation results coincide
with the theoretical analysis.

\begin{figure}[h]
\centerline{
\includegraphics[width=1.8in,height=1.5in]{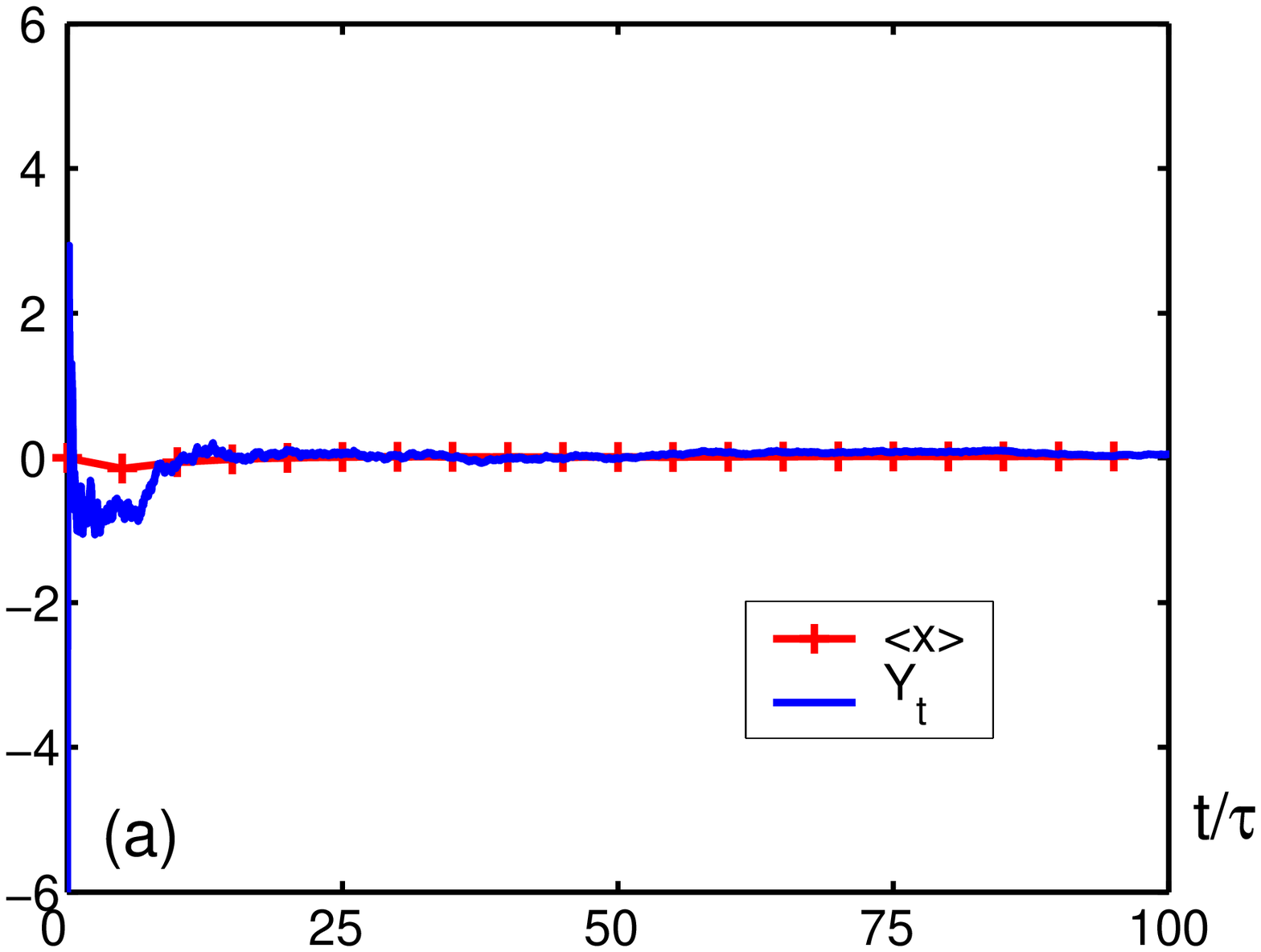}
\includegraphics[width=1.8in,height=1.5in]{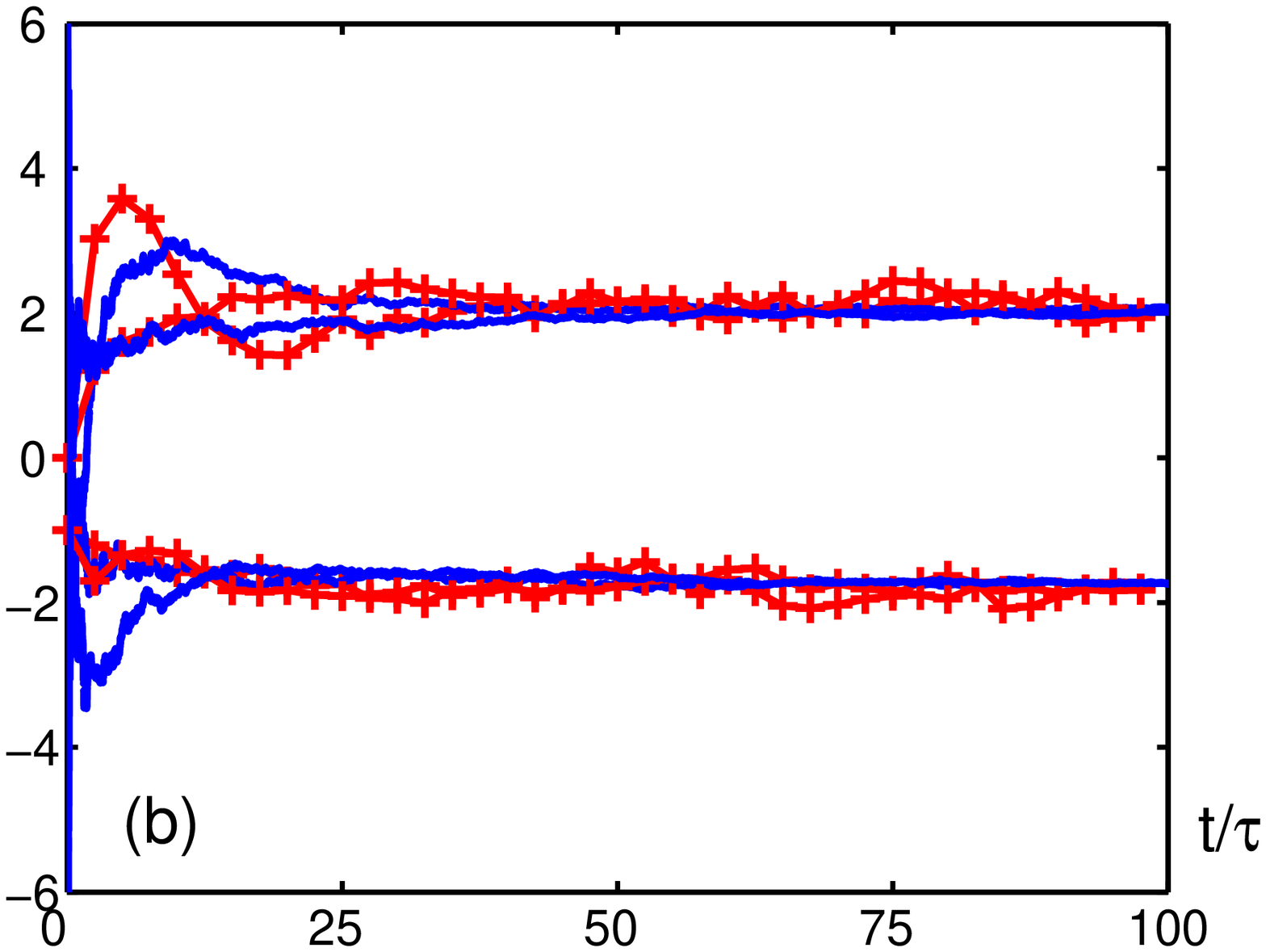}
} \caption{(Color online) Stochastic pitchfork bifurcation given
by Eqs.~(\ref{General stochastic master equation}), (\ref{General
measurement output}), and (\ref{Three-order nonlinear feedback
control}) with (a) $\omega=3.5$ MHz and (b) $\omega=65$ MHz. The
parameter $\tau=2$ ns is the normalization time scale. The red
curves with plus signs (one curve in Fig.~\ref{Fig of the
pitchfork bifurcation}a and four curves in Fig.~\ref{Fig of the
pitchfork bifurcation}b) represent the stochastic trajectories
$\langle x \rangle_t$, and the blue solid curves (one curve in
Fig.~\ref{Fig of the pitchfork bifurcation}a and four curves in
Fig.~\ref{Fig of the pitchfork bifurcation}b) denote the
time-average trajectories $Y_t$ given by Eq.~(\ref{Time-averaged
measurement output}).}\label{Fig of the pitchfork bifurcation}
\end{figure}

As analyzed in Appendix \ref{Dynamics of the Controlled Harmonic
Oscillator}, the first-order quadratures $\langle x\rangle$ and
$\langle p\rangle$ of the controlled system evolve nonlinearly in
the long-time limit, while the higher-order quadratures approach
to those of the linear quantum systems which preserve the Gaussian
properties of the states. Therefore, the nonlinear dynamics
induced by the proposed quantum feedback control is
``semiclassical" in some sense (see the simplified diagram of the
feedback control circuit in Fig.~\ref{Fig of the feedback control
circuit}), in contrast to the dynamics of the system governed by a
fully quantum nonlinear Hamiltonian (about $x$ and $p$) such as:
\begin{equation}\label{Four-order nonlinear Hamiltonian}
H_{\rm nl}=\frac{1}{2} p^2+\frac{1}{2}\omega^2 x^2-k x^4,
\end{equation}
in which the Gaussian wavepackets are distorted. Here the
subscript ``nl" means that the Hamiltonian contains higher-order
nonlinear terms. In our feedback control proposal, we introduce
nonlinear terms like $\langle x \rangle ^3$ to the dynamical
equation of the system in comparison to the terms $\langle x ^3
\rangle$ introduced by the nonlinear Hamiltonian $H_{\rm nl}$ in
Eq.~(\ref{Four-order nonlinear Hamiltonian}). Thus, under the
nonlinear Hamiltonian $H_{\rm nl}$, the equations of the first
order quadratures $\langle x \rangle$, $\langle p \rangle$ and the
second order quadratures $V_x$, $V_p$ and $C_{xp}$ are not closed,
since higher-order quadratures such as $\langle x^3 \rangle$ are
involved. However, in our proposal, the equations of $\langle x
\rangle$, $\langle p \rangle$ and $V_x$, $V_p$, $C_{xp}$ are
closed and decoupled just like linear harmonic oscillator if the
initial state of the harmonic oscillator is a Gaussian state (see
the analysis in the Appendix~\ref{Dynamics of the Controlled
Harmonic Oscillator}), whose stationary solution can be
analytically solved. For this reason, we call this
feedback-control-induced nonlinear dynamics as ``pseudo" nonlinear
dynamics. As shown below, interesting phenomena can be observed
when such a ``pseudo" nonlinear system is coupled to another
quantum system. That is, similar to the nonlinear amplifier using
the nonlinear dynamical bifurcation (e.g., rf-driven Josephson
junctions~\cite{Siddiqi1}), the harmonic oscillator with feedback
control can be used to amplify small signals, and furthermore read
out the qubit states.
\begin{figure}[h]
\includegraphics[bb=79 326 453 533, width=7.5 cm, clip]{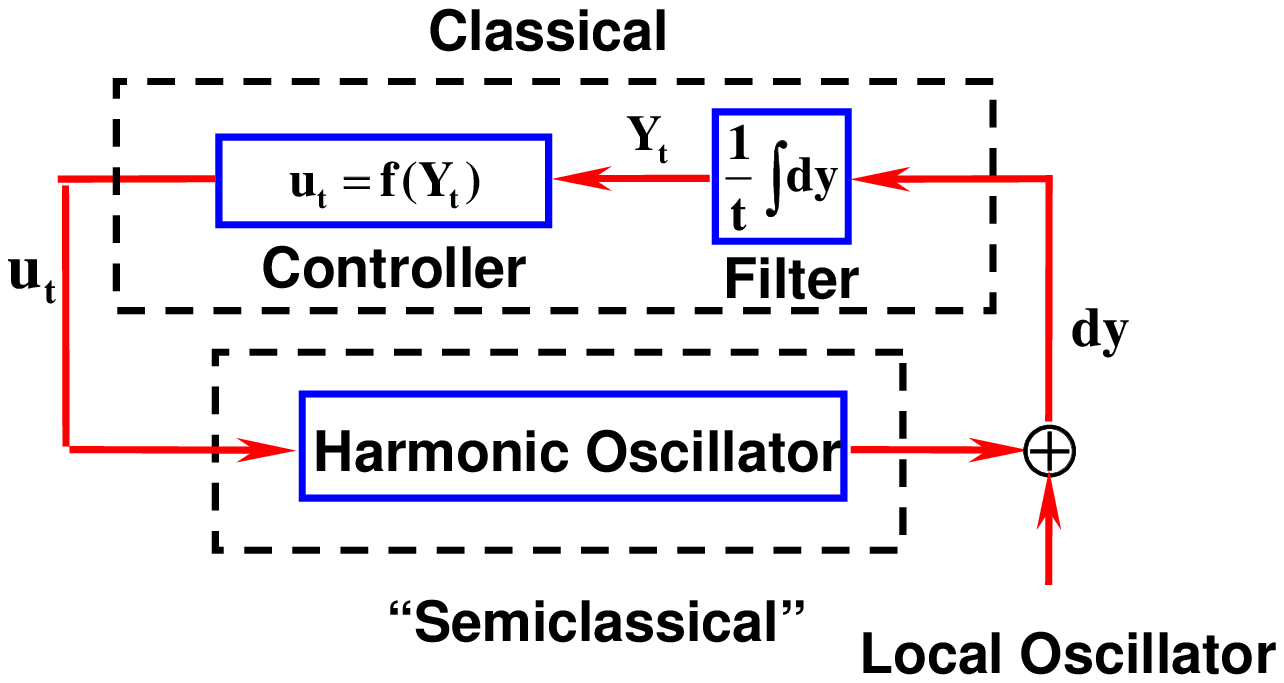}
\caption{(Color online) Schematic diagram of the feedback control
circuit: the harmonic oscillator is measured by a homodyne
detection, and the output signal of the homodyne measurement is
fed into a classical control circuit which is composed of a
integral filter and a controller. The output signal of the control
circuit is further fed back to drive the harmonic oscillator, and
this feedback control signal may induce nonlinear dynamics of the
harmonic oscillator.}\label{Fig of the feedback control circuit}
\end{figure}

\section{Measurements on qubit states using controlled harmonic
oscillator}\label{s3}

\subsection{Qubit-oscillator coupling}\label{s31}

To demonstrate how we can use the ``pseudo" nonlinear harmonic
oscillator produced by feedback control to measure qubit states,
let us consider a coupled qubit-oscillator system driven by a
classical field with angular frequency $\omega_d$. The Hamiltonian
of this system can be expressed as:
\begin{eqnarray}\label{Hamiltonian of the qubit-resonator system}
H&=&\frac{\hbar\omega_q}{2}\sigma_z+\hbar\omega_o
a^{\dagger}a+\hbar
g(a^{\dagger}\sigma_-+a\sigma_+)\nonumber\\
&&+\hbar u_t\frac{1}{\sqrt{2}}\left( a^{\dagger}e^{-i\omega_d t}+
a e^{i\omega_d t}\right),
\end{eqnarray}
where $\omega_q$ and $\omega_o$ are the angular frequencies of the
qubit and the harmonic oscillator; $\sigma_z$ and $\sigma_{\pm}$
are the $z$-axis Pauli operator and ladder operators of the qubit;
$a$ and $a^{\dagger}$ are the annihilation and creation operators
of the harmonic oscillator; $g$ is the qubit-oscillator coupling
strength; and $u_t$ is the coupling constant between the driving
field and the harmonic oscillator.

In the dispersive regime, i.e.,
$|\Delta_{qo}|=|\omega_q-\omega_o|\gg|g|$, we can apply the
following unitary transformation
\begin{eqnarray*}
U=\exp\left[\frac{g}{\Delta_{qo}}\left(a\sigma^{\dagger}-a^{\dagger}\sigma_-\right)\right]
\end{eqnarray*}
to the Hamiltonian $H$ in Eq.~(\ref{Hamiltonian of the
qubit-resonator system}), then we have an effective Hamiltonian
\begin{eqnarray}
\tilde{H}=UHU^{\dagger}&\approx&\frac{\hbar\left(\omega_q+\chi\right)}{2}\sigma_z+\hbar\omega_o
a^{\dagger}a+\hbar\chi a^{\dagger}a\sigma_z\nonumber\\
&&+\hbar
u_t\frac{1}{\sqrt{2}}\left(a^{\dagger}e^{-i\omega_d\,t}+ae^{i\omega_d\,t}\right),\label{eq:6}
\end{eqnarray}
where $\chi=g^2/\Delta_{qo}$ is the effective coupling strength
between the qubit and the harmonic oscillator.

Furthermore, the Hamiltonian $\tilde{H}$ can be rewritten in the
rotating reference frame under the unitary transformation
\begin{eqnarray*}
U_{\rm rot}=\exp\left[i\left(\omega_d t\right)a^{\dagger}a\right],
\end{eqnarray*}
thus we can obtain the following effective Hamiltonian
\begin{eqnarray}\label{Effective Hamitonian of the qubit-resonator system in the large-detuning regime}
H_{\rm eff}&=&U_{\rm rot}\tilde{H}U_{\rm
rot}^{\dagger}+i\dot{U}_{\rm rot}U_{\rm rot}^{\dagger}\nonumber\\
&=&\frac{\hbar\omega_q}{2}\sigma_z+\hbar\Delta_{od}
a^{\dagger}a+\hbar\chi a^{\dagger}a\sigma_z+\hbar u_t x,
\end{eqnarray}
where $\Delta_{od}=\omega_o-\omega_d$ is the angular frequency
detuning between the harmonic oscillator and the driving field;
and $x=\left(a+a^{\dagger}\right)/\sqrt{2}$ is the position
operator of the harmonic oscillator. Here the frequency-shift of
the qubit caused by the harmonic oscillator has been neglected
under the assumption that $\omega_q\gg\chi$ which is usually valid
in atom-optical systems and superconducting circuits.

If we expand the Hamiltonian in Eq.~(\ref{eq:6}) to
$\left(g/\Delta_{qo}\right)^3$ terms, we can obtain the following
effective nonlinear Hamiltonain:
\begin{eqnarray}\label{Effective Hamiltonian under four-order perturbation expansion}
\tilde{H}_{\rm
eff}&=&\frac{\hbar}{2}\left(\omega_q+\frac{g^2}{\Delta_{qo}}+\frac{g^4}{2\Delta_{qo}^3}\right)\sigma_z+\hbar\Delta_{od} a^{\dagger}a\nonumber\\
&&+\hbar\frac{g^2}{\Delta_{qo}}a^{\dagger}a\sigma_z-\hbar\frac{g^4}{\Delta_{qo}^3}\left(a^{\dagger}a\right)^2\sigma_z+\hbar
u_t x.
\end{eqnarray}
The last second term in Eq.~(\ref{Effective Hamiltonian under
four-order perturbation expansion}) may induce nonlinear dynamics
to the harmonic oscillator. However,
$\left(g/\Delta_{qo}\right)^2$ times smaller compared with the
third term in Eq.~(\ref{Effective Hamiltonian under four-order
perturbation expansion}), it is too small to be observed in the
large-detuning regime $g\ll\Delta_{qo}$. Thus, the high-order
terms have been neglected in below discussions.

The decay of the harmonic oscillator is detected by a homodyne
measurement. Thus, the evolution of the qubit-oscillator system is
conditioned on the measurement output of the homodyne detection,
which can be described by the stochastic master equation
(\ref{General stochastic master equation}) by replacing the system
Hamiltonian $H$ with $H_{\rm eff}$ given in Eq.~(\ref{Effective
Hamitonian of the qubit-resonator system in the large-detuning
regime}). The output signal of the homodyne detection is
integrated to obtain a new output signal $Y_t=\int_0^t dy/t$,
which is further fed into a nonlinear controller to produce the
control signal $u_t$ given in Eq.~(\ref{Three-order nonlinear
feedback control}). As presented in Sec.~\ref{s2}, such a simple
third-order nonlinear feedback control induces a static
bifurcation that can be used to enhance the measurement strength
for the qubit readout.

Different from the open-loop control that is predetermined by the
designer without any information extraction, the proposed feedback
control $u_t$ is automatically adjusted according to the state of
the qubit in real time so that different controls generate
different state trajectories and thus different output signals.
This makes it possible to identify the state of the qubit by
amplifying the difference between the output signals by the
designed control. This feature of feedback control has been
reported in the literature to enhance the measurement intensity by
linear amplification (see, e.g., Ref.~\cite{Combes}), which can be
done more efficiently via nonlinear amplification induced by the
proposed quantum feedback control.

Let us assume that the qubit-oscillator system is initially in a
separable state
\begin{eqnarray*}
\rho(0)=\rho_q(0)\otimes|\psi_0(0)\rangle\langle\psi_0(0)|,
\end{eqnarray*}
where
\begin{eqnarray*}
\rho_q(0)=\sum_{i,j=g,e}\rho_{ij}(0)|i\rangle\langle j|
\end{eqnarray*}
is the initial state of the qubit with ground state $|g\rangle$
and excited state $|e\rangle$; and $|\psi_0(0)\rangle$ is a
Gaussian state of the harmonic oscillator with the first and
second-order quadratures given in Eq.~(\ref{First and second-order
quadratures of the initial Gaussian state}). Then, the stationary
state of the qubit-oscillator system in the long-time limit can be
expressed as~\cite{Gambetta}:
\begin{equation}\label{Stationary state of the qubit-resonator system}
\rho^{\infty}=\sum_{i,j=g,e}\rho_{ij}^{\infty}|i\rangle\langle
j|\otimes|\alpha_i^{\infty}\rangle\langle\alpha_j^{\infty}|,
\end{equation}
where both $|\alpha_e^{\infty}\rangle$ and
$|\alpha_g^{\infty}\rangle$ are coherent states. Let
\begin{eqnarray*}
\alpha_{g,e}^{\infty}=\frac{1}{\sqrt{2}}\left(x_{g,e}^{\infty}+ip_{g,e}^{\infty}\right),
\end{eqnarray*}
then $\left(x_g^{\infty},p_g^{\infty}\right)^T$ and
$\left(x_e^{\infty},p_e^{\infty}\right)^T$ are respectively the
stationary states of the equations:
\begin{eqnarray}\label{Centroid evolution corresponding to two eigen states of charge qubit}
\dot{x}_{g,e}&=&-\frac{\gamma}{2}x_{g,e}+\left(\Delta_{od}\mp\chi\right)p_{g,e},\\
\dot{p}_{g,e}&=&-\frac{\gamma}{2}p_{g,e}-\left(\Delta_{od}\mp\chi-k_1\right)x_{g,e}-k_3
x_{g,e}^3+k_0.\nonumber
\end{eqnarray}
The coefficients $\rho_{ij}^{\infty}$ in Eq.~(\ref{Stationary
state of the qubit-resonator system}) are given by:
\begin{eqnarray*}
&\rho_{gg}^{\infty}=\rho_{gg}(0),\quad\rho_{ee}^{\infty}=\rho_{ee}(0),\quad\rho_{ge}^{\infty}=\rho_{eg}^{\infty}=0.&
\end{eqnarray*}

As analyzed in the last paragraph of Sec.~\ref{s2}, the harmonic
oscillator can be looked as a ``semiclassical" pseudo nonlinear
system driven by an outer classical feedback control circuit (see
the simplified version of the feedback control circuit in
Fig.~\ref{Fig of the feedback control circuit of the coupled
qubit-resonator system}). The interaction between such a pseudo
nonlinear system and the qubit brings two aspects of effects. On
the one hand, for the qubit, this interaction brings additional
decoherence. In fact, it can be found that there exists a
measurement-induced dephasing factor for the reduced states of the
qubit of which the damping rate can be approximately estimated in
the long-time limit as (see the analysis in Appendix
\ref{Calculations of the measurement-induced damping rate}):
\begin{equation}\label{Measurement-induced dephasing factor}
\Gamma=\chi\left(x_e^{\infty}p_g^{\infty}-p_e^{\infty}x_g^{\infty}\right).
\end{equation}
On the other hand, this interaction leads to additional frequency
shift for the harmonic oscillator depending on the state of the
qubit. Therefore it is possible to dispersively read out the
states of the qubit under appropriate conditions.
\begin{figure}[h]
\includegraphics[bb=78 319 396 546, width=7.5 cm, clip]{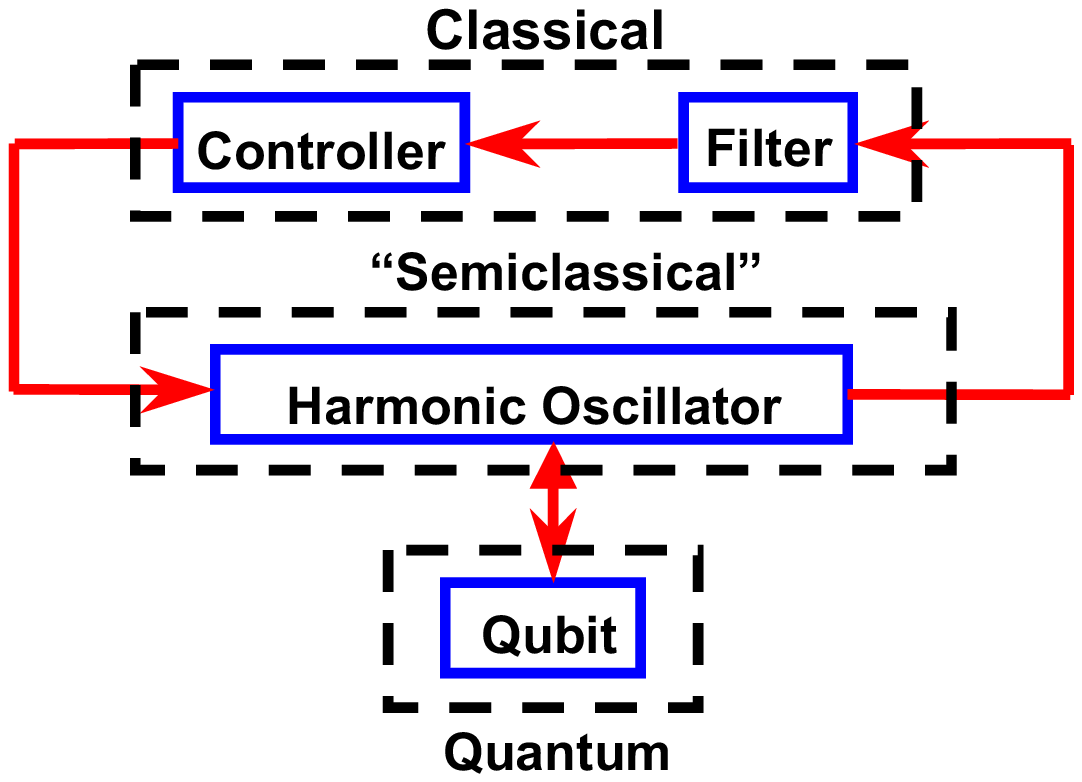}
\caption{(Color online) Simplified version for the feedback
control circuit of the coupled qubit-oscillator system. The qubit
is coupled to the harmonic oscillator, which is driven by a
feedback control circuit composed of a filter and a controller.
The nonlinear feedback control coming from the control circuit
leads to ``semiclassical" nonlinear dynamics of the harmonic
oscillator. The static bifurcation induced by such nonlinear
dynamics can be used to amplify small signals and thus measure the
states of the qubit.}\label{Fig of the feedback control circuit of
the coupled qubit-resonator system}
\end{figure}

\subsection{Bifurcation-induced quantum measurement}\label{s31}

The equations of the harmonic oscillator corresponding to the
ground and excited states of the qubit given in Eq.~(\ref{Centroid
evolution corresponding to two eigen states of charge qubit}) have
the same form as in Eq.~(\ref{Dynamical equation with three-order
nonlinear terms}) with two different angular frequencies
$\omega=\Delta_{od}-\chi$ and $\omega=\Delta_{od}+\chi$. As
analyzed in Sec.~\ref{s2}, a static pitchfork bifurcation occurs
at a critical angular frequency
\begin{eqnarray*}
\omega^*=\frac{1}{2}\left(k_1-\sqrt{k_1^2-\gamma^2}\right).
\end{eqnarray*}
When $\omega<\omega^*$, the system given by Eq.~(\ref{Dynamical
equation with three-order nonlinear terms}) possesses only one
stable equilibrium, which becomes unstable when $\omega>\omega^*$
and, in the meanwhile, two other stable equilibria appear.

Therefore, by tuning the angular frequency $\omega_d$ of the
driving field near the bifurcation point, the static bifurcation
introduced by the proposed feedback control can induce a
transition from weak to strong measurements for the qubit readout.
Actually, if
\begin{equation}\label{Driving angular frequency without bifurcation}
\omega_d=\omega_o-\omega^*+2\chi,
\end{equation}
both effective angular frequencies $\omega_g=\omega^*-3\chi$ and
$\omega_e=\omega^*-\chi$ corresponding to the two eigenstates of
the qubit are lower than $\omega^*$. Then, in the long-time limit,
e.g., when $t\gg1/\gamma$, the measurement outputs corresponding
to the two eigenstates of the qubit are
\begin{equation}\label{Measurement outputs for linear regime}
Y_{g,e}(t)\rightarrow
x_{g,e}^{\infty}=k_0\frac{\omega_{g,e}}{\omega_{g,e}^2-k_1\omega_{g,e}+\gamma^2/4}
\end{equation}
respectively (see Eq.~(\ref{x0infinity}) and the analysis in
Appendix \ref{Dynamics of the Controlled Harmonic Oscillator}). If
$k_0$ is small enough such that
\begin{equation}\label{k0}
k_0\ll\left|\frac{\chi^2-2\chi\omega^*+k_1\chi}{\omega^*-\chi}\right|,
\end{equation}
both $Y_g(t)$ and $Y_e(t)$ are so close to zero that they are
almost indistinguishable. When $t\gg1/\gamma$, we have
\begin{eqnarray*}
x_{g,e}&\rightarrow&x_{g,e}^{\infty}=\frac{k_0\omega_{g,e}}{\omega_{g,e}^2-k_1\omega_{g,e}+\gamma^2/4},\\
p_{g,e}&\rightarrow&p_{g,e}^{\infty}=\frac{\gamma}{2\omega_{g,e}}x_{g,e}^{\infty},
\end{eqnarray*}
then we can calculate the measurement-induce dephasing rate
$\Gamma_{\rm weak}$ from Eq.~(\ref{Measurement-induced dephasing
factor}):
\begin{eqnarray}\label{Gammaweak}
\Gamma_{\rm
weak}&=&\chi\left(x_e^{\infty}p_g^{\infty}-p_e^{\infty}x_g^{\infty}\right)\nonumber\\
&=&\frac{\gamma
k_0^2\chi^2}{\left[\omega_g^2-k_1\omega_g+\frac{\gamma^2}{4}\right]\left[\omega_e^2-k_1\omega_e+\frac{\gamma^2}{4}\right]}.\nonumber\\
\end{eqnarray}

The tiny separation between $Y_g(t)$ and $Y_e(t)$ can be amplified
by switching the angular frequency of the driving field from
$\omega_d=\omega_o-\omega^*+2\chi$ to
$\tilde{\omega}_d=\omega_o-\omega^*$.
Under this condition, the two effective angular frequencies
$\tilde{\omega}_g=\omega^*-\chi$ and
$\tilde{\omega}_e=\omega^*+\chi$ are lower and higher than the
bifurcation angular frequency $\omega^*$ respectively. When
$t\gg1/\gamma$, the corresponding stationary measurement outputs
are
\begin{eqnarray}\label{Measurement outputs at the bifurcation point}
Y_g(t)&\rightarrow&
\tilde{x}_g^{\infty}=k_0\frac{\tilde{\omega}_g}{\tilde{\omega}_g^2-k_1\tilde{\omega}_g+\gamma^2/4},\nonumber\\
Y_e(t)&\rightarrow&\tilde{x}_e^{\infty}=\sqrt{\frac{-\tilde{\omega}_e^2+k_1\tilde{\omega}_e
-\gamma^2/4}{k_3\tilde{\omega}_e}}.
\end{eqnarray}
By setting
\begin{equation}\label{k3}
k_3\ll\left|\frac{\chi^2+2\chi\omega^*-k_1\chi}{\omega^*+\chi}\right|,
\end{equation}
we have
\begin{eqnarray*}
|\tilde{x}_e^{\infty}-\tilde{x}_g^{\infty}|\gg|x_e^{\infty}-x_g^{\infty}|.
\end{eqnarray*}
The above result means that the difference between the two output
signals corresponding to the two eigenstates of the qubit is
amplified by the proposed nonlinear feedback control. When
$t\gg1/\gamma$, it can be calculated that
\begin{eqnarray*}
x_g&\rightarrow&\tilde{x}_g^{\infty}=k_0\frac{\tilde{\omega}_g}{\tilde{\omega}_g^2-k_1\tilde{\omega}_g+\gamma^2/4},\\
x_e&\rightarrow&\tilde{x}_e^{\infty}=\sqrt{\frac{-\tilde{\omega}_e^2+k_1\tilde{\omega}_e
-\gamma^2/4}{k_3\tilde{\omega}_e}},\\
p_g&\rightarrow&\tilde{p}_g^{\infty}=\frac{\gamma}{2\tilde{\omega}_g}\tilde{x}_g^{\infty},\\
p_e&\rightarrow&\tilde{p}_e^{\infty}=\frac{\gamma}{2\tilde{\omega}_e}\tilde{x}_e^{\infty}.
\end{eqnarray*}
Then, we can obtain the measurement-induced dephasing rate
$\Gamma_{\rm strong}$ from Eq.~(\ref{Measurement-induced dephasing
factor}) as:
\begin{eqnarray}\label{Gammastrong}
\Gamma_{\rm
strong}&=&\chi\left(\tilde{x}_e^{\infty}\tilde{p}_g^{\infty}-\tilde{p}_e^{\infty}\tilde{x}_g^{\infty}\right)\nonumber\\
&=&\frac{k_0\gamma\chi^2}{\sqrt{k_3\tilde{\omega}_e^3}}\cdot\frac{\sqrt{-\tilde{\omega}_e^2+k_1\tilde{\omega}_e-\gamma^2/4}}{\tilde{\omega}_g^2-k_1\tilde{\omega}_g+\gamma^2/4}.
\end{eqnarray}
It can be noticed that $\Gamma_{\rm strong}$ is far greater than
$\Gamma_{\rm weak}$ when $k_0$ and $k_3$ are sufficiently small to
satisfy Eqs.~(\ref{k0}) and (\ref{k3}).

\section{Applications}\label{s5}
\subsection{Atom-optical systems}\label{s51}
Consider a coupled atom-cavity system, in which a two-level $C_s$
atom is dispersively coupled to a single-mode field in an optical
cavity~\cite{Hood1} (see Fig.~\ref{Fig of the atom-optical system
with feedback control loop}). The probe light transmitted through
the optical cavity is detected by a homodyne measurement. Then,
the output photoncurrent is fed into a control circuit composed of
an integral filter and a controller to generate a control signal,
which is further fed back to control the probe light by a
modulator.
\begin{figure}[h]
\centerline{
\includegraphics[width=3.2in,height=2.3in]{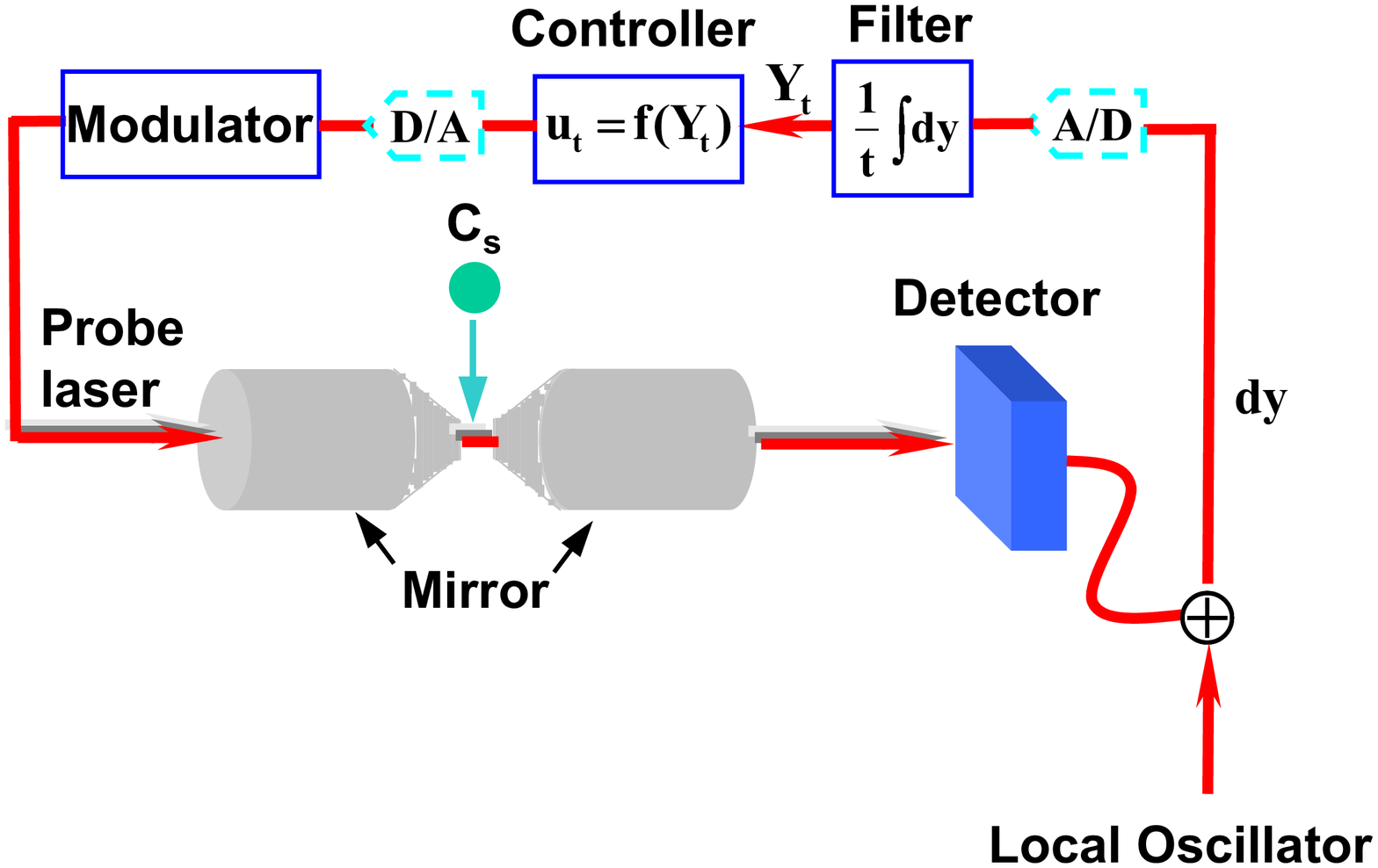}
} \caption{(Color online) Schematic diagram of the atom-optical
system under feedback control. A two-level $C_s$ atom is coupled
to a single-mode quantum field in the optical cavity, whose output
field is detected by a homodyne measurement (``LO" denotes a local
oscillator). The output signal of the homodyne measurement is fed
into an electric control circuit to generate the desired control
signal, which is further fed back to modulate the probe laser fed
into the optical cavity.  The output signal of the homodyne
detection may also be converted into a digital signal via an
analog/digital (A/D) signal convertor, and thus the control signal
can be generated by a Digital Signal Processor (DSP), which is
further converted into an electric signal via a digital/analog
(D/A) signal convertor and fed back.}\label{Fig of the
atom-optical system with feedback control loop}
\end{figure}

We choose the following experimentally accessible parameters (see,
e.g., Refs.~\cite{Hood1,Hood2,Puppe}):
\begin{eqnarray*}
\Delta_{qo}/2\pi&=&35\,\,{\rm MHz},\\
g/2\pi&=&8\,\,{\rm
MHz},\\
\gamma/2\pi&=&1.4\,\,{\rm MHz},
\end{eqnarray*}
where $\Delta_{qo}=\omega_q-\omega_o$ is the detuning between the
frequency $\omega_q$ of the two-level atom and the frequency
$\omega_o$ of the single-mode field in the optical cavity; $g$ is
the coupling constant between the two-level atom and the optical
cavity; and $\gamma$ is the damping rate of the optical cavity.
The control parameters $k_0$, $k_1$, and $k_3$ are chosen to be
\begin{eqnarray*}
k_0/2\pi&=&1\,\,{\rm MHz},\\
k_1/2\pi&=&6\,\,{\rm MHz},\\
k_3/2\pi&=&1\,\,{\rm MHz}.
\end{eqnarray*}
We first tune the detuning $\Delta_{od}=\omega_o-\omega_d$ between
the frequency $\omega_o$ of the optical cavity and the frequency
$\omega_d$ of the driving field such that
\begin{eqnarray*}
\Delta_{od}/2\pi=-3.57\,\,{\rm MHz}.
\end{eqnarray*}
In this case, the long-time limits of the measurement outputs
$Y_g$ and $Y_e$ corresponding to the ground and excited states
$|g\rangle$ and $|e\rangle$ of the two-level atom are both close
to zero, and thus almost indistinguishable (see Fig.~\ref{Fig of
the bifurcation-induced measurement for atom-optical systems}a).
This corresponds to the weak measurement case.

If we tune the detuning frequency $\Delta_{od}$ such that
\begin{eqnarray*}
\Delta_{od}/2\pi=0.083\,\,{\rm MHz},
\end{eqnarray*}
the two branches of outputs corresponding to $|g\rangle$ and
$|e\rangle$ are separated in the long-time limit, which
corresponds to the strong measurement case (see Fig.~\ref{Fig of
the bifurcation-induced measurement for atom-optical systems}b).
\begin{figure}[h]
\centerline{
\includegraphics[width=1.8in,height=1.3in]{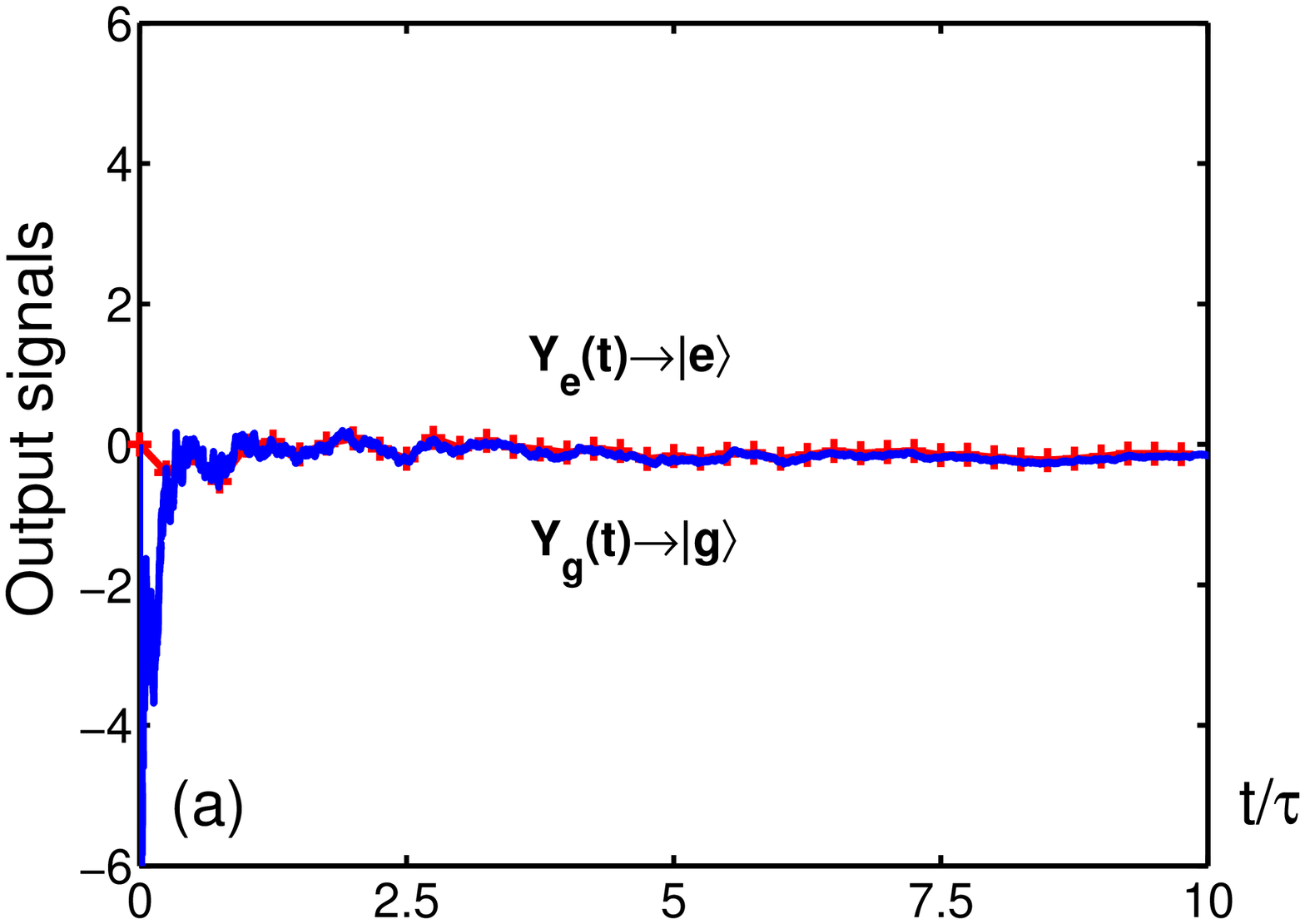}
\includegraphics[width=1.8in,height=1.3in]{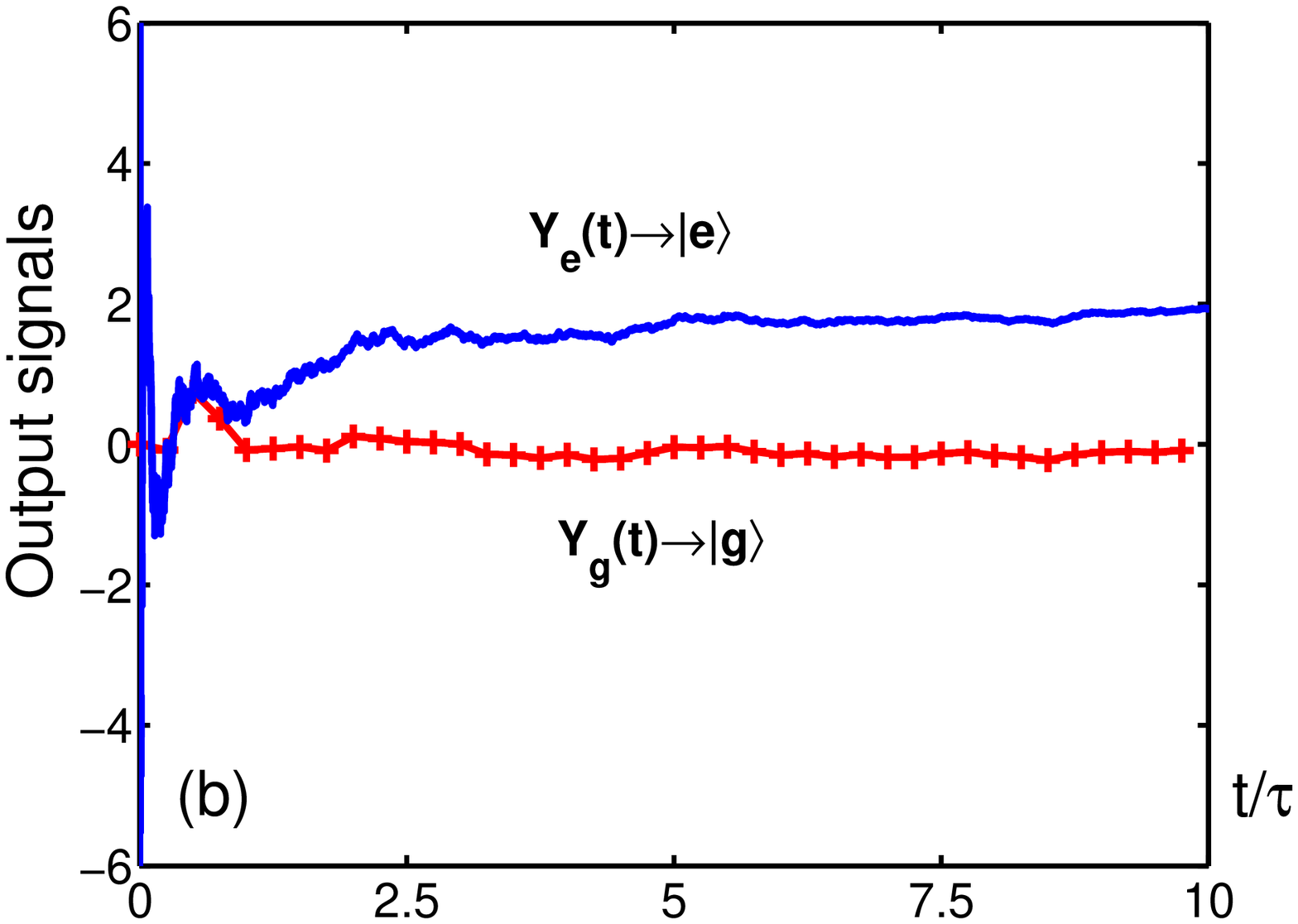}
} \caption{(Color online) Measurement outputs of the atom-cavity
system under feedback control. (a) $\Delta_{od}/2\pi=-3.57$ MHz
(weak measurement case): the two output signals $Y_g(t)$ (red
curve with plus signs) and $Y_e(t)$ (blue solid curve)
corresponding to the two eigenstates $|g\rangle$ and $|e\rangle$
of the two-level atom are almost indistinguishable; (b)
$\Delta_{od}/2\pi=0.083$ MHz (strong measurement case): the two
output signals $Y_g(t)$ (red curve with plus signs) and $Y_e(t)$
(blue solid curve) are separated in the long time limit. The
parameter $\tau=0.16$ $\mu$s is the normalization time
scale.}\label{Fig of the bifurcation-induced measurement for
atom-optical systems}
\end{figure}
\subsection{Superconducting circuits}\label{s52}
The second example considers the superconducting circuit shown in
Fig.~\ref{Fig of the superconducting circuit with feedback control
loop}, in which a transmission line resonator is capacitively
coupled to a Cooper pair box (charge qubit).
\begin{figure}[h]
\centerline{
\includegraphics[width=3in,height=2.7in]{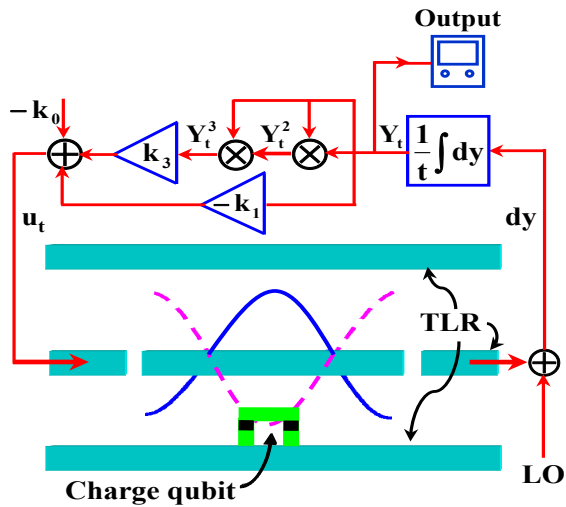}
} \caption{(Color online) Schematic diagram of the superconducting
circuit under feedback control. A charge qubit is capacitively
coupled to a transmission line resonator (TLR), whose output field
is detected by a homodyne measurement (``LO" denotes a local
oscillator). The output signal of the homodyne measurement is fed
into an electric control circuit to generate the desired control
signal, which is further used to drive the electric field in the
transmission line resonator.}\label{Fig of the superconducting
circuit with feedback control loop}
\end{figure}
Let us now discuss experimental feasibility via the numerical
simulation using the experimentally accessible parameters.
According to the current experiments (see, e.g.,
Ref.~\cite{Circuit QED}), the qubit frequency $\omega_{q}$, the
frequency $\omega_o$ and the decay rate $\gamma$ of the
transmission line resonator, as well as the coupling constant $g$
between the qubit and the transmission line resonator can be
chosen as:
\begin{eqnarray}\label{System parameter}
&\omega_q/2\pi =5.1\,\,{\rm GHz}, \quad \omega_o/2\pi=5\,\,{\rm
GHz},&\nonumber\\
&\gamma/2\pi=100\,\,{\rm MHz}, \quad g/2\pi=20\,\,{\rm MHz}.&
\end{eqnarray}
With the conditions given in Eqs. (\ref{k0}) and (\ref{k3}), we
further assume that the parameters $k_{0}$, $k_{1}$, and $k_{3}$
are
\begin{eqnarray}
k_0/2\pi&=&20\,\,{\rm MHz},\nonumber\\
k_1/2\pi&=&200\,\,{\rm MHz},\nonumber\\
k_3/2\pi&=&2\,\,{\rm MHz}\,\,{\rm or}\,\,10\,\,{\rm MHz}.
\end{eqnarray}
The frequency $\omega_d/2\pi$ of the driving field is initially
chosen to be $4.995$ GHz. At the time $t^*/\tau=100$, i.e.,
$t^*=50$ ns, the frequency $\omega_d/2\pi$ of the driving field is
switched from $4.995$ GHz to $4.987$ GHz, where $\tau=0.5$ ns is a
normalization time scale. Simulation results in Fig.~\ref{Fig of
the bifurcation-induced measurement} show that, at time $t^*$, the
difference between the two output signals $Y_g(t)$ and $Y_e(t)$,
i.e., the measurement sensitivity, suddenly jumps. The
measurement-induced dephasing rate also suddenly jumps from
$\Gamma_{\rm weak}$ to $\Gamma_{\rm strong}$. In fact, it can be
calculated from Eqs.~(\ref{Gammaweak}) and (\ref{Gammastrong})
that $\Gamma_{\rm weak}/2\pi\approx0.36$ MHz and $\Gamma_{\rm
strong}/2\pi\approx10.22$ MHz (or $4.57$ MHz) when $k_3/2\pi=2$
MHz (or $10$ MHz). This indicates that the static bifurcation
introduced by our proposal induces a transition from weak to
strong measurements at time $t^*$. Moreover, as shown in
Eq.~(\ref{Measurement outputs at the bifurcation point}) and the
simulation results in Fig.~\ref{Fig of the bifurcation-induced
measurement}, the decrease of the nonlinear coefficient $k_3$
makes the measurement more sensitive, but accelerates the
dephasing of the qubit. {\it Such a tradeoff between measurement
sensitivity and measurement-induced dephasing effects is a natural
consequence of the confliction between information extraction and
measurement-induced disturbance, which is inherent for quantum
measurement.}
\begin{figure}[h]
\centerline{
\includegraphics[width=2.6in,height=1.8in]{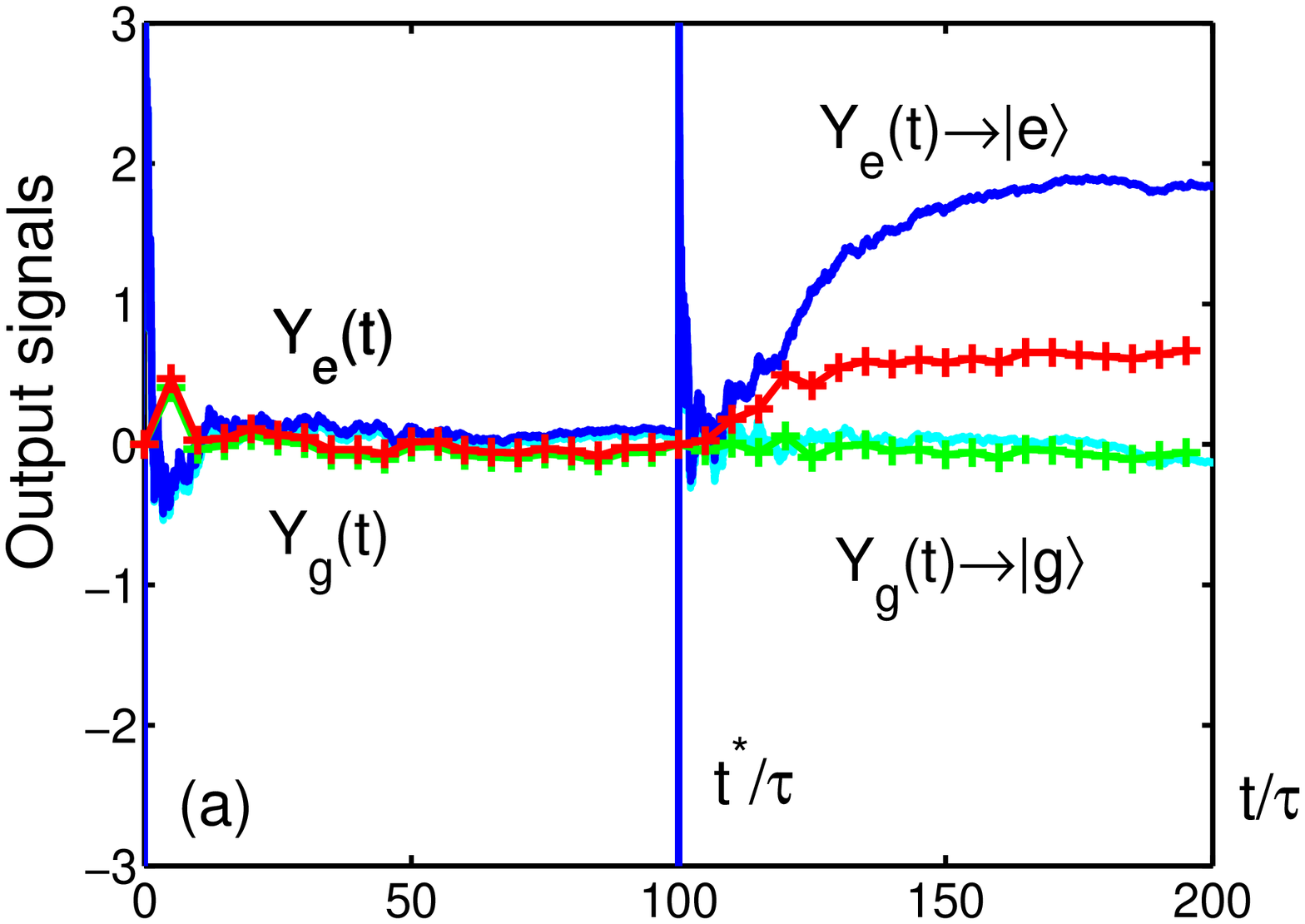}}
\centerline{
\includegraphics[width=2.6in,height=1.8in]{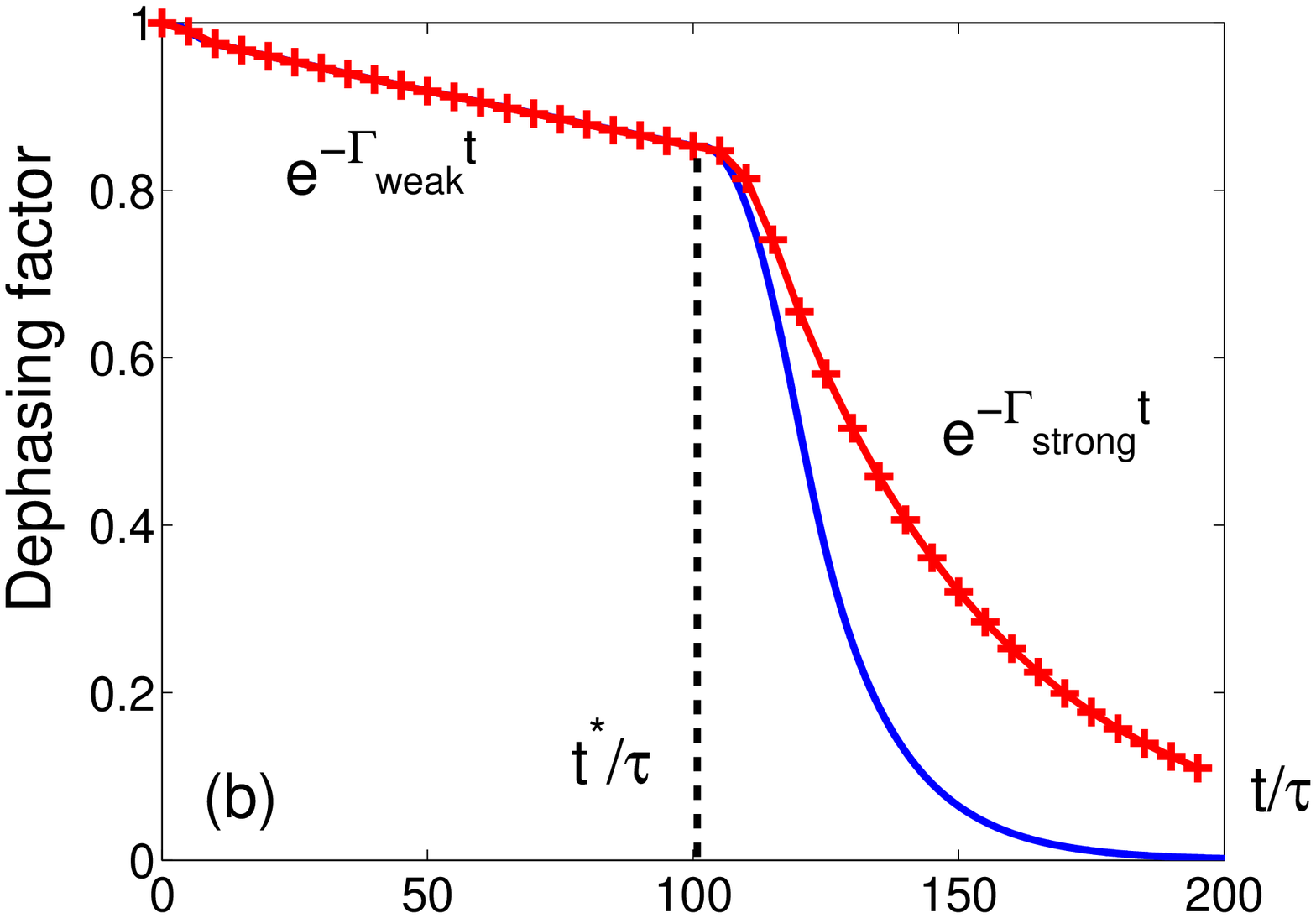}
} \caption{(Color online) Bifurcation-induced transition from a
weak measurement to a strong measurement with (a) the two output
signals $Y_g(t)$ and $Y_e(t)$ corresponding to the two eigenstates
$|g\rangle$ and $|e\rangle$ of the charge qubit, and (b) the
measurement-induced dephasing factor. The parameter $\tau=0.5$ ns
is the normalization time scale. The solid curves represent the
trajectories with a nonlinear coefficient $k_3/2\pi=2$ MHz, while
the solid curves with plus signs represent the trajectories with
$k_3/2\pi=10$ MHz.}\label{Fig of the bifurcation-induced
measurement}
\end{figure}

\section{Conclusion}\label{s6}

In summary, we present a method to induce ``semiclassical"
nonlinear dynamics in a damped harmonic oscillator via quantum
feedback control. The nonlinear feedback control induces a static
bifurcation, which can be used to amplify the small
frequency-shift of the harmonic oscillator for the qubit readout.
Theoretical analysis and numerical simulations show an evident
transition from weak to strong measurements near the bifurcation
point. Our proposal works as well as the bifurcation readout
proposal by a nonlinear amplifier, e.g., a rf-driven Josephson
junction. Additionally, we can tune the information extraction
rate to balance between information extraction and
measurement-induced disturbance in the ``nonlinear" readout
regime. Using the atom-optical systems and circuit-QED systems as
examples, we show how to apply our proposal to experimental
systems.

We emphasize that the proposed cavity-assistant nonlinear
amplification strategy can also be applied to the qubit readouts
in other experimental implementations, such as quantum dot-cavity
systems. This generalized result is important, in that the
proposed measurement method might be more efficient, because the
nonlinear amplification device, like the rf-driven Josephson
junctions in superconducting circuits, may not be achievable.

Our study is also hopeful to be extended to quantum demolition
measurements, e.g., the detection of momentum or position of a
nanomechanical resonator, for which the measurement-induced
back-action effects on the quantity being measured cannot be
neglected.

\begin{center}
\textbf{ACKNOWLEDGMENTS}
\end{center}

We acknowledge financial support from the National Natural Science
Foundation of China under Grant Nos. 60704017, 10975080, 60904034,
60836001, 60635040. T.~J. Tarn would also like to acknowledge
partial support from the U. S. Army Research Office under Grant
W911NF-04-1-0386.
\\[0.2cm]

\appendix
\section{Dynamics of the Controlled Harmonic
Oscillator}\label{Dynamics of the Controlled Harmonic Oscillator}
From the stochastic master equation (\ref{General stochastic
master equation}) and the measurement output equation
(\ref{General measurement output}) given in Sec.~\ref{s2}, we can
obtain the following non-Markovian stochastic integro-differential
equations for $\langle x \rangle$ and $\langle p \rangle$:
\begin{eqnarray}\label{Integro-differential equations for the first-order quadratures}
d\langle x \rangle&=&-\frac{\gamma}{2}\langle x \rangle
dt+\omega\langle p \rangle
dt+\sqrt{2\eta\gamma}\left(V_x-\frac{1}{2}\right),\nonumber\\
d\langle p \rangle&=&-\omega\langle x \rangle
dt-\frac{\gamma}{2}\langle p \rangle dt+k_0
dt+\sqrt{2\eta\gamma}C_{xp}dW\nonumber\\
&&+k_1\frac{1}{t}\int_0^t\left(\langle x \rangle
dt+\frac{1}{\sqrt{2\eta\gamma}}dW\right)\nonumber\\
&&-k_3\left[\frac{1}{t}\int_0^t\left(\langle x \rangle
dt+\frac{1}{\sqrt{2\eta\gamma}}dW\right)\right]^3,
\end{eqnarray}
which are equivalent to stochastic Markovian equations by
introducing a new variable $Y_t$:
\begin{eqnarray}\label{Dynamical equations for the first-order quadratures}
d\langle x\rangle&=&-\frac{\gamma}{2}\langle x\rangle
dt+\omega\langle p\rangle
dt+\sqrt{2\eta\gamma}\left(V_x-\frac{1}{2}\right)dW,\nonumber\\
d\langle p\rangle&=&-\omega\langle x\rangle
dt-\frac{\gamma}{2}\langle p\rangle dt+\left(k_1Y_t-k_3Y_t^3+k_0\right)dt\nonumber\\
&&+\sqrt{2\eta\gamma}C_{x p}dW,\nonumber\\
dY_t&=&-\frac{1}{t}\left(Y_t-\langle x
\rangle\right)dt-\frac{1}{t\sqrt{2\eta\gamma}}dW,
\end{eqnarray}
where
\begin{eqnarray*}
V_x=\langle x^2 \rangle-\langle x \rangle^2,\quad V_p=\langle p^2
\rangle-\langle p \rangle^2
\end{eqnarray*}
are the variances of the position $x$ and momentum $p$; and
\begin{eqnarray*}
C_{xp}&=&\langle\frac{xp+px}{2}\rangle-\langle x \rangle\langle p
\rangle
\end{eqnarray*}
is the symmetric covariance of $x$ and $p$.

Taking the average of the above equations, we have
\begin{eqnarray}\label{Average dynamical equation for the first-order quadratures}
\dot{\bar{x}}&=&-\frac{\gamma}{2}\bar{x}+\omega\bar{p},\nonumber\\
\dot{\bar{p}}&=&-\omega\bar{x}-\frac{\gamma}{2}\bar{p}+k_1\bar{Y}_t-k_3\bar{Y}_t^3+k_0,\nonumber\\
\dot{\bar{Y}}_t&=&-\frac{1}{t}\left(\bar{Y}_t-\bar{x}\right),
\end{eqnarray}
where $\bar{x}=E\left(\langle x \rangle\right)$,
$\bar{p}=E\left(\langle p \rangle\right)$, and
$\bar{Y}_t=E\left(Y_t\right)$. Here, we neglect the high-order
correlation of $Y_t$ induced by the classical noise $dW$ which is
reasonable when the evolution time is sufficiently long.

When $\omega<\omega^*$, we have
\begin{eqnarray}\label{vec(x)_0}
x_0^{\infty}&=&k_0\frac{\omega}{\omega^2-k_1\omega+\gamma^2/4}>0,\nonumber\\
p_0^{\infty}&=&k_0\frac{\gamma/2}{\omega^2-k_1\omega+\gamma^2/4}>0.
\end{eqnarray}
Then, since we start from the initial state
$\left(0,0,0\right)^T$, it can be verified from Eq.~(\ref{Average
dynamical equation for the first-order quadratures}) that
\begin{eqnarray*}
0\leq\bar{x}\leq x_0^{\infty},\quad 0\leq\bar{p}\leq
p_0^{\infty},\quad 0\leq\bar{Y}_t\leq x_0^{\infty}.
\end{eqnarray*}
 Thus, from the last equation in Eq.~(\ref{Average dynamical
equation for the first-order quadratures}), we have
$\dot{\bar{Y}}_t \rightarrow 0$, i.e., $\exists Y^{\infty}$, s.t.,
$\bar{Y}_t\rightarrow Y^{\infty}$ when $t\rightarrow \infty$.
Substituting $Y^{\infty}$ into the first two equations of
Eq.~(\ref{Average dynamical equation for the first-order
quadratures}), we have
\begin{eqnarray*}
\left(%
\begin{array}{c}
  \dot{\bar{x}} \\
  \dot{\bar{p}} \\
\end{array}%
\right)&=&\left(%
\begin{array}{cc}
  -\frac{\gamma}{2} & \omega \\
  -\omega & -\frac{\gamma}{2} \\
\end{array}%
\right)\left(%
\begin{array}{c}
  \bar{x} \\
  \bar{p} \\
\end{array}%
\right)\\
&&+\left(%
\begin{array}{c}
  0 \\
  k_1 Y^{\infty}-k_3 Y^{\infty\,3}+k_0 \\
\end{array}%
\right).
\end{eqnarray*}
Since the matrix
\begin{eqnarray*}
\left(%
\begin{array}{cc}
  -\frac{\gamma}{2} & \omega \\
  -\omega & -\frac{\gamma}{2} \\
\end{array}%
\right)<0,
\end{eqnarray*}
we know that there exists $\left(x^{\infty},p^{\infty}\right)^T$
such that
\begin{eqnarray*}
\left(%
\begin{array}{c}
  \bar{x}(t) \\
  \bar{p}(t) \\
\end{array}%
\right)\rightarrow \left(%
\begin{array}{c}
  x^{\infty} \\
  p^{\infty} \\
\end{array}%
\right),
\end{eqnarray*}
when $t\rightarrow\infty$.

Next, we will show that $Y^{\infty}=x^{\infty}$. Note that
\begin{eqnarray*}
\bar{Y}_t=E\left(Y_t\right)=\frac{1}{t}\int_0^t\bar{x}\left( \tau
\right)d\tau,
\end{eqnarray*}
we know that $\forall\,\epsilon>0$, $\exists\,T_{\sigma}$, s.t.
$\forall\,t>T_{\sigma}$,
$\left|\bar{x}-x^{\infty}\right|<\epsilon/2$. Let
\begin{eqnarray*}
\Delta(T_{\sigma})=\max_{\left[0,T_{\sigma}\right]}\left|\bar{x}-x^{\infty}\right|
\end{eqnarray*}
and
\begin{eqnarray*}
T_Y=\max\left\{T_{\sigma},2\epsilon^{-1}T_{\sigma}\Delta(T_{\sigma})\right\}.
\end{eqnarray*}
Then, $\forall\,t>T_Y$, we have
\begin{eqnarray*}
\left|\bar{Y}_t-x^{\infty}\right|&<&\frac{1}{t}\int\limits_0^{T_{\sigma}}\left|\bar{x}(s)-x^{\infty}\right|ds\\
&&+\frac{1}{t}\int\limits_{T_{\sigma}}^t\left|\bar{x}(s)-x^{\infty}\right|ds\\
&<&\frac{\epsilon}{2T_{\sigma}\Delta(T_{\sigma})}\cdot
T_{\sigma}\Delta(T_{\sigma})+\frac{\epsilon}{2}=\epsilon,
\end{eqnarray*}
which means that $Y^{\infty}=x^{\infty}$.

In order to solve $\left(x^{\infty},p^{\infty}\right)^T$, we let
$\dot{\bar{x}}=0$, $\dot{\bar{p}}=0$ to obtain the algebraic
equations
\begin{eqnarray*}
0&=&-\frac{\gamma}{2}x^{\infty}+\omega p^{\infty},\\
0&=&-\omega x^{\infty}-\frac{\gamma}{2}p^{\infty}+k_1
x^{\infty}-k_3 \left(x^{\infty}\right)^3+k_0.
\end{eqnarray*}
Although three equilibria can be solved in this case, only one
equilibrium $\left( x_0^{\infty}, p_0^{\infty} \right)^T$ is
stable when $\omega<\omega^*$, where $x_0^{\infty}$,
$p_0^{\infty}$ are given by Eq.~(\ref{vec(x)_0}). Thus, if
$\omega<\omega^*$, we have
\begin{eqnarray}\label{Long-time limit of the first-order quadratures}
E\left(\langle x \rangle\right)&=&\bar{x}\rightarrow
x_0^{\infty},\nonumber\\
E\left( \langle p \rangle \right) & = & \bar{p}\rightarrow
p_0^{\infty},\nonumber\\
E\left( Y_t \right) & = & \bar{Y}_t \rightarrow
Y_0^{\infty}=x_0^{\infty},
\end{eqnarray}
when $t\rightarrow\infty$.

Furthermore, it can be calculated that the equations of
higher-order quadratures are independent of the control $u_t$.
Thus, the evolutions of higher-order quadratures are like those of
linear harmonic oscillator. Since we start from a Gaussian state,
the state given by Eq.~(\ref{General stochastic master equation})
remains to be a Gaussian state, and thus we only need to calculate
the variances $V_x,\,V_p$ and the symmetric covariance $C_{xp}$.
From Eq.~(\ref{General stochastic master equation}), we can obtain
that
\begin{eqnarray}\label{Dynamical equation for the second-order quadratures}
\dot{V}_x&=&-\gamma V_x+2\omega C_{x
p}+\frac{\gamma}{2}-2\eta\gamma\left(V_x-\frac{1}{2}\right)^2,\nonumber\\
\dot{V}_p&=&-\gamma V_p-2\omega C_{x
p}+\frac{\gamma}{2}-2\eta\gamma C_{x p}^2,\nonumber\\
\dot{C}_{x p}&=&-\gamma C_{x p}+\omega V_p-\omega
V_x-2\eta\gamma\left(V_x-\frac{1}{2}\right)C_{x p},\nonumber\\
\end{eqnarray}
from which it can be verified that
\begin{eqnarray}\label{Long-time limit of the second-order quadratures}
V_x&\rightarrow&V_x^{\infty}=1/2,\nonumber\\
V_p&\rightarrow&V_p^{\infty}=1/2,\nonumber\\
C_{xp}&\rightarrow&C_{xp}^{\infty}=0,
\end{eqnarray}
when $t\rightarrow\infty$. From Eqs.~(\ref{Long-time limit of the
first-order quadratures}) and (\ref{Long-time limit of the
second-order quadratures}), we know that the state of the
controlled harmonic oscillator tends to a stationary coherent
state $|\alpha_0^{\infty}\rangle$ given by
Eq.~(\ref{alpha0infinity}) when $\omega<\omega^*$.

With the same discussions, it can be verified that there exist two
stable equilibria for Eq.~(\ref{Average dynamical equation for the
first-order quadratures}) when $\omega>\omega^*$ which can be
expressed as:
\begin{equation}\label{vec(x12)}
\left(%
\begin{array}{c}
  x_{1,2}^{\infty} \\
  p_{1,2}^{\infty} \\
  Y_{1,2}^{\infty} \\
\end{array}%
\right)=\pm\sqrt{\frac{-\omega^2+k_1\omega-\gamma^2/4}{k_3\omega}}\left(%
\begin{array}{c}
  1 \\
  2\omega/\gamma \\
  1 \\
\end{array}%
\right),
\end{equation}
and Eq.~(\ref{Long-time limit of the second-order quadratures})
can also be obtained in this case. It means that the state of the
controlled harmonic oscillator tends to
$|\alpha_1^{\infty}\rangle$ or $|\alpha_2^{\infty}\rangle$ when
$\omega>\omega^*$.

\section{Approximate estimation of the measurement-induced
dephasing rate}\label{Calculations of the measurement-induced
damping rate}

With the same analysis as in Appendix A of Ref.~\cite{Gambetta2},
the state of the qubit-oscillator system can be expressed as:
\begin{eqnarray*}
\rho(t)=\sum_{i,j=g,e}\rho_{ij}\left(t\right)\left|i\rangle\langle
j\right|\otimes\left|\psi_i(t)\rangle\langle\psi_j(t)\right|,
\end{eqnarray*}
where $|\psi_e\rangle$ and $|\psi_g\rangle$ are Gaussian states
with first and second-order quadratures in the phase space as:
\begin{eqnarray*}
&\langle x \rangle_{g,e}=x_{g,e},\quad \langle p
\rangle_{g,e}=p_{g,e},&\\
&\langle x^2 \rangle_{g,e}-\langle x \rangle^2_{g,e}=V_{g,e}^x,&\\
&\langle p^2 \rangle_{g,e}-\langle p \rangle^2_{g,e}=V_{g,e}^p,&\\
&\left\langle\frac{xp+px}{2}\right\rangle_{g,e}-\langle x
\rangle_{g,e}\langle p \rangle_{g,e}=C_{g,e}^{xp}.&
\end{eqnarray*}
The first-order quadratures $x_{g,e}$, $p_{g,e}$ satisfy the
equations:
\begin{eqnarray*}
d\,x_{g,e}&=&-\frac{\gamma}{2}x_{g,e}dt+\left(\Delta_{od}\mp\chi\right)p_{g,e}dt\\
&&+\sqrt{2\eta\gamma}\left(V_{g,e}^x-\frac{1}{2}\right)dW,\\
d\,p_{g,e}&=&-\left(\Delta_{od}\mp\chi\right)x_{g,e}dt-\frac{\gamma}{2}p_{g,e}dt\\
&&+\left(k_1Y_{g,e}-k_3
Y_{g,e}^3+k_0\right)dt+\sqrt{2\eta\gamma}C_{g,e}^{ep}dW,\\
d\,Y_{g,e}&=&-\frac{1}{t}\left(Y_{g,e}-x_{g,e}\right)dt-\frac{1}{t\sqrt{2\eta\gamma}}dW,
\end{eqnarray*}
and the second-order quadratures $V_{g,e}^x$, $V_{g,e}^p$,
$C_{g,e}^{xp}$ satisfy the equations:
\begin{eqnarray*}
\dot{V}_{g,e}^x&=&-\gamma
V_{g,e}^x+2\left(\Delta_{od}\mp\chi\right)C_{g,e}^{xp}+\frac{\gamma}{2}\\
&&-2\eta\gamma\left(V_{g,e}^x-\frac{1}{2}\right)^2,\\
\dot{V}_{g,e}^p&=&-\gamma
V_{g,e}^p-2\left(\Delta_{od}\mp\chi\right)C_{g,e}^{xp}+\frac{\gamma}{2}-2\eta\gamma\left(C_{g,e}^{xp}\right)^2,\\
\dot{C}_{g,e}^{xp}&=&-\gamma
C_{g,e}^{xp}+\left(\Delta_{od}\mp\chi\right)V_{g,e}^p-\left(\Delta_{od}\mp\chi\right)V_{g,e}^x\\
&&-2\eta\gamma\left(V_{g,e}^x-\frac{1}{2}\right)C_{g,e}^{xp}.
\end{eqnarray*}
Here, as an approximate estimation, we omit the higher-order
disturbance induced by $V_{g,e}^x$, $V_{g,e}^p$, and
$C_{g,e}^{xp}$, then the coefficients $\rho_{ij}(t)$ are given by
(see, e.g., Ref.~\cite{Gambetta}):
\begin{eqnarray*}
&\rho_{gg}(t)=\rho_{gg}(0),\quad\rho_{ee}(t)=\rho_{ee}(0),\quad\rho_{ge}(t)=\rho_{eg}^*(t),&\\
&\rho_{eg}(t)=\frac{\exp\left[-\left(\gamma_2
t+\Sigma(t)\right)-i\left(\omega_q
t+\Theta(t)\right)\right]}{\langle\psi_g(t)|\psi_e(t)\rangle}\rho_{eg}(0),&
\end{eqnarray*}
where $\gamma_2$ is the dephasing rate of the qubit without
measurement; and $\Sigma(t)$, $\Theta(t)$ can be expressed as:
\begin{eqnarray}
\label{Sigma(t)}\Sigma(t)&=&\chi\int_0^t\left[x_e(s)p_g(s)-p_e(s)x_g(s)\right]ds,\\
\label{Theta(t)}\Theta(t)&=&\chi\int_0^t\left[x_e(s)x_g(s)+p_e(s)p_g(s)\right]ds.
\end{eqnarray}
By tracing out the degrees of freedom of the harmonic oscillator,
it can be shown that there exists a measurement-induced dephasing
factor $\exp\left[-\Sigma(t)\right]$.

As is proved in Appendix \ref{Dynamics of the Controlled Harmonic
Oscillator}, we have $x_{g,e}\rightarrow x_{g,e}^{\infty}$,
$p_{g,e}\rightarrow p_{g,e}^{\infty}$ in the long-time limit.
Thus, for any $s>\bar{t}\gg1/\gamma$, it can be approximately
estimated as $x_{g,e}(s)\approx x_{g,e}^{\infty}$,
$p_{g,e}(s)\approx p_{g,e}^{\infty}$. Then, the
measurement-induced dephasing after $\bar{t}$ can be approximately
calculated as:
\begin{eqnarray*}
&&\exp\left\{\chi\int_{\bar{t}}^t\left[x_e(s)p_g(s)-p_e(s)x_g(s)\right]ds\right\}\\
&\approx&\exp\left\{\left[\chi\left(x_e^{\infty}p_g^{\infty}-p_e^{\infty}x_g^{\infty}\right)\right]\left(t-\bar{t}\right)\right\},
\end{eqnarray*}
which leads to Eq.~(\ref{Measurement-induced dephasing factor}).
\\[0.2cm]

\end{document}